\documentclass[epj]{svjour}
\usepackage{graphics}
\usepackage{amssymb,amsmath}
\newcommand{\be}{\begin{equation}}
\newcommand{\ee}{\end{equation}}
\newcommand{\ba}{\begin{eqnarray}}
\newcommand{\ea}{\end{eqnarray}}
\newcommand{\n}{\nonumber}
\newcommand{\lan}{\langle}
\newcommand{\ran}{\rangle}
\newcommand{\A}{\alpha}
\newcommand{\B}{\beta}

\newcommand{\sn}{\langle n \rangle}
\newcommand{\ket}[1]{\bigl|#1\bigr\rangle}
\newcommand{\bra}[1]{\bigl\langle #1\bigr|}
\newcommand{\ketbra}[2]{\bigl|#1\bigr\rangle\bigl\langle #2\bigr|}

\newcommand{\Tr}[2][]{\mathrm{Tr}_{#1}\left\{ #2 \right\}}
\newcommand{\adj}{^{\dagger}}
\newcommand{\expect}[1]{\bigl\langle #1\bigr\rangle}

\begin{document}

\title{Violation of Bell's inequality for continuous variables }
\author{L. Praxmeyer\inst{1} \and B.-G.~Englert\inst{2} \and
K. W\'odkiewicz\inst{1}$^{,}$\inst{3} 
}                     

\institute{Institute of Theoretical Physics, Warsaw University,
Ho\.za 69, 00--681 Warsaw, Poland  \and Department of Physics,
National University of Singapore, Singapore 117542\and Department
of Physics and Astronomy, University of New Mexico,
 Albuquerque, NM 87131, USA}
\date{Received: date / Revised version: date}
%
\abstract{
We construct a wide class of bounded  continuous variables
observables that lead to violations of Bell inequalities for the
EPR state and  give an intuitive Wigner function explanation how
to predetermine which operators won't ever exceed the bounds given by
local theories.
} 

 \maketitle

\section{Introduction}
\label{intro} Bell's inequality was derived and tested for
entangled system of two qubits (polarization or spin) \cite{photon,aspect,weihs,10km,ion,atom}. Recent
investigations have been dealing with systems described by
continuous variables (CV) \cite{kluw} such as the original
 Einstein-Podolsky-Rosen (EPR) example or entangled pairs of photons
generated in non degenerate optical parametric amplification
(NOPA).
A simple way of implementing Bell measurements on CV systems is to
use dichotomic  (bounded by $\pm 1$) observables. Recent examples
of such observables are the parity operator \cite{BW2,BW3}
 or CV spin operators \cite{chen}.
It is the purpose of this work to give a wide class of quantum
observables that can be implemented into correlation measurements
of entangled states. We show that such bounded operators will
often have quite different properties in the Wigner representation
-- the representation that provides a fundamental link between
classical and quantum physics. The Wigner function gives a natural
phase-space framework in which the relation between local realism
and quantum probability rules can be formulated and studied. The
original EPR wave function is a Gaussian state with a nonnegative
Wigner function, which can be interpreted as a  hidden phase-space
probability distribution. As we have already mentioned, the main
goal of this paper is to construct a wide class of such bounded
continuous variables observables that will lead to violations of
Bell inequalities for the EPR state.

The paper is organized as follows: Sections \ref{sec2} and
\ref{sec3} provide an introduction to the Wigner representation of
quantum correlations, its connection with local theories and the
CV form of the EPR state. In Section \ref{sec4} we introduce a
class of bounded observables and choose from them certain
representatives that realize the Pauli algebra.   The physical
interpretation of these operators is presented in Section
\ref{sec5}. The final Section \ref{sec6} shows explicitly that
some of these operators lead to a violation of Bell's inequality.
Results presented there were obtained either analytically or by
rather simple  numerics. Concluding remarks are offered in Section
\ref{sum}.

\section{Entanglement in the Wigner representation}
\label{sec2}

 As an example let us probe a
two-party entangled  system, described by a non separable density
operator $\rho$, for correlations. The probing  of the
entanglement can be achieved by a joint measurement performed by
Alice and Bob using local observables $A$ and $B$. In this
measurement, Alice and Bob measure  a correlation  $\lan A,B\ran =
\Tr{ (A\otimes B)\rho}$. Using Wigner functions we can write this
quantum correlation as \be \label{corr} \lan A,B\ran = \int
(d\lambda_a) (d\lambda_b) W_{A}(\lambda_a)\, W_{B}(\lambda_b)\,
W_{\rho} (\lambda_a, \lambda_b) \,, \ee where $(d\lambda_{a})$,
$(d\lambda_{b})$ are properly normalized measures of the phase
space variables $(q_{a},p_{a})$ and $(q_{b},p_{b})$, respectively.
The three Wigner functions correspond to the observables $A$,
$B$  (associated with Alice and Bob) and to the entangled state
$\rho$. This formula has a remarkable structure of a local hidden
variable theory if one entertains the association that $W_{A}$,
$W_{B}$ correspond to ``hidden" predetermined values of operators
$A,\,B$, and $ W_{\rho} (\lambda_a, \lambda_b)$ is a genuine local
probability distribution function of the hidden variables.

From these correlation we can form the following Bell combination:
\be \label{bell1} {\mathcal{B}} = \lan A,B\ran + \lan A, B'\ran +
\lan A',B\ran- \lan A', B'\ran\,. \ee In the local hidden
variables theory Bell's inequality $ |{\mathcal{B}}|\leq 2 $
should be satisfied if these observables fulfill the boundary
conditions $|A|\leq 1$  and $|B|\leq 1$. For systems described by
continuous variables, the selection of these variables is not as
obvious as in the case of measurements performed on entangled
qubits. This inequality can be treated as a test dividing purely
quantum phenomena from those that can by explained by
deterministic models. A violation of the Bell's inequality
(\ref{bell1}) means that the effect we study requires a quantum
description.

\section{The EPR state}
\label{sec3}
As an example of an entangled state leading to quantum CV
correlations (\ref{corr}), we use a two-mode squeezed state. It is
well known that the CV form of the EPR  state can be generated in
a non degenerate optical parametric amplification involving two
modes of the radiation field \cite{kimble,kimble2}. The wave
function  of such a pure quantum state $\rho =
\ketbra{\Psi}{\Psi}$ has the Schmidt decomposition

\ba
|\Psi\ran&=\frac{1}{\cosh r}\sum_{n=0}^{\infty} (\tanh r)^n|n,n\ran=\n\\
&=\frac{1}{\sqrt{1+\sn}}\sum_{n=0}^{\infty}
\bigg(\frac{\sn}{1+\sn}\bigg)^{\frac{n}{2}}|n,n\ran\,,
\label{NOPA} \ea where $\sn$ is the mean number of photons in each
mode and $r$ denotes the squeezing parameter. These two
parameterizations are connected by the relation $\sn= \sinh^2 r$.
 In the
limit of $\sn \rightarrow \infty$, the two-mode squeezed state
becomes the original EPR state. The Wigner function of this state
is given by
\ba
&& 4\pi^2\, W_{\Psi}(q_a,p_a,q_b,p_b)\label{wig_nopa}\\
&&=\exp{\bigl(-(1+2\sn)(p_{a}^2+p_{b}^2)-4\sqrt{\sn(\sn+1)}p_a
p_b \bigr)}\n\\
&&\times\exp{\bigl(-(1+2\sn)(q_{a}^2+q_{b}^2)+4\sqrt{\sn(\sn+1)}q_a q_b
\bigr)}\,.\n
\ea
As it has been mentioned in the Introduction, the nonnegative
 Wigner function of the EPR state can be interpreted as a
 probability distribution of CV local realities.

It is worth  noticing that this is a unique case when the Wigner
function is exactly  equal  to the product of the probabilities in
the position and momentum representations,

\be W_{\Psi}(q_a,p_a,q_b,p_b)=\frac{1}{(2\pi)^2} |\tilde{\Psi}
(p_a,p_b)|^2|\Psi (q_a,q_b)|^2\label{wig}. \ee

The factors are, of course, the marginal probabilities as obtained
from the wave functions implied by Eq. (\ref{NOPA}),

 \ba &&\Psi
(q_a,q_b)=\frac{1}{\sqrt{\pi}}
e^{-(\sn+\frac{1}{2})(q_a^2+{q_b}^2) + 2\sqrt{\sn(1+\sn)} q_a
q_b}\n,\\
 \label{nopa3}
&&\widetilde{\Psi} (p_a,p_b) =\sqrt{\pi}
e^{-(\sn+\frac{1}{2})(p_a^2+{p_b}^2) - 2\sqrt{\sn(1+\sn)} p_a
p_b}\n, \ea
a consequence of the fact that the state considered is
a Gaussian state with no position-momentum correlation between the
two particles. We stress that it is really an untypical situation
and the Wigner function here has an even more intuitive
interpretation than usually.

The Wigner function of the entangled CV state can be used to
describe the correlations between massive particles formed in a breakup
process, or for clouds of cold atoms.

\section{Bounded observables}
\label{sec4}
Our goal is to construct quantum observables  for Alice and Bob that are
bounded by $\pm 1$ . For Alice we introduce  a class of quantum
observables of the
form
\be
 A = \int dq \, a_{\epsilon} (q) \ketbra{q}{\epsilon
q}\, \ee in the position representation, where $\epsilon \neq 0$
is a real parameter and $a_{\epsilon}(q) $ is a function of $q$.
In the same way one can construct quantum observables for Bob.

This operator is hermitian ($A= A\adj$) if $\epsilon=\pm 1$ and
 \be
a_{\epsilon}( q)=  a^{\ast}_{\epsilon} (\frac{q}{\epsilon})\,. \ee
In this case we  have $a_{+}(q)= a^{\ast}_{+}(q)$ or $a_{-}(q)=
a_{-}^{\ast}(-q)$. The condition that this observable has a sharp
bound, $A^2=1$, is satisfied if
$a_{\epsilon}( q)a_{\epsilon}(\epsilon q)=1\,.$

The Wigner functions of these dichotomic operators with $\epsilon=\pm 1$ are
\ba W_{A_{a_{+}}}(q,p) &= & \frac{1}{2\pi} a_{+}(q)\,,
\nonumber\\
W_{A_{a_{-}}}(q,p) &= & \frac{1}{2}\delta(q) \int \frac{d\xi}{2\pi}
e^{ip\xi}\, a_{-}(\frac{\xi}{2})\,,
 \ea
where we recognize in the last expression the Fourier transform of
$a_{-}(q)$. In the case of $\epsilon=-1$, the corresponding Wigner
function is never bounded, leading to a possible violation of the
Bell inequalities. The simplest example of such an observable $A$
is the parity operator $\mathbb{P}$, \ba && \mathbb{P}= \int dq
\ketbra{q}{-q}, \ea corresponding to $a_{-}(q)=1$. The Wigner
function of this observable is \be
W_{\mathbb{P}}(q,p)=\frac{1}{2}\delta(q)\delta(p) . \ee This
dichotomic operator has been used recently to probe Bell
inequalities for systems described by continuous variables
\cite{BW2,BW3}.

Another simple example of a dichotomic operator is $a_{+}(q) =
\mathrm{sgn}(q)$, corresponding to the sign operator $\mathbb{S}$,
\be \mathbb{S}= \int dq \, \mathrm{sgn}(q)\ketbra{q}{q}\,. \ee The
corresponding Wigner function \be W_{\mathbb{S}}(q,p)=
\frac{1}{2\pi} \mathrm{sgn}(q) \ee is bounded and no violation of
Bell inequalities should be expected. This example shows that the
quantum nonlocality of the EPR state cannot be revealed by
measuring  quadrature components.

As another  example, let us consider a function $a_{-}(q) = i\,
\mathrm{sgn}(q)$. This function
 defines a hermitian
operator that we shall call the parity inversion,
\be
\mathbb{R} =i \int dq \, \mathrm{sgn}(q)\ketbra{q}{-q}\,.
\ee
The Wigner function of this observable is unbounded:
\be
W_{\mathbb{R}}(q,p)=-\frac{1}{2}\delta(q)\mathcal{P}\frac{1}{p}\,,
\ee
($\mathcal{P}$ denotes the Cauchy principal value). Certainly this
singular and unbounded function can be used to exhibit the
nonlocality of the EPR state.

The three hermitian operators that we have introduced satisfy the
commutation relations for the Pauli matrices,
\be [\mathbb{S}, \mathbb{R}] =2 i \mathbb{P}, \ [\mathbb{P},
\mathbb{S}] = 2i \mathbb{R}, \
 [\mathbb{R}, \mathbb{P}] = 2i \mathbb{S}\,. \ee
 Different in form representations of the commutation relations
 presented above have been given in the recent literature
\cite{chen,fiurasek,gour}.

It is well known that the phase-space shift of an observable can be
implemented with the help of  the local  displacement operator
$D(q,p)$ that is familiar from the theory of coherent states.
In the
following section we will  use such  shifts in order to form the
Bell combination. 
For Alice and Bob we introduce
shifted operators $A(\alpha)= D(\alpha) A D\adj(\alpha)$ and
$B(\beta)= D(\beta) A D\adj(\beta)$, where the two complex numbers
$\alpha$ and $\beta$ characterize the phase space shifts in Alice's
and Bob's position and momentum $(q,p)$. These parameters are the CV
analogues of the polarization settings for qubits.

For an unsharp bound of the observables, the condition for
$\epsilon$ is less restrictive. We will give examples of such
unsharp functions at the end of this paper.

\section{Measurements by Alice and Bob}
\label{sec5}
 The expectation values of
$\mathbb{P}$, $\mathbb{R}$, $\mathbb{S}$ provide their physical
interpretation (or at least operational meaning associated with
position measurements):
{\small \ba
&&\expect{\mathbb{P}} = \int \,dq\,\Psi^*(q) \Psi(-q)\,,  \n\\
&&\expect{\mathbb{S}} =
\int_{R_+}\,dq \,|\Psi(q)|^2 -\int_{R_-}\,dq\, |\Psi(q)|^2 \,,
\label{srednie}\\
&&\expect{\mathbb{R}} = i \biggl( \int_{R_+}\, dq\,
\Psi^*(q)\Psi(-q) -\int_{R_-}\, dq\,\Psi^*(q)\Psi(-q)\biggr). \n
\ea } According to the equations above the expectation value of
$\mathbb{P}$ is (up to the normalization factor) equal to the
Wigner function value at the origin, $W(0,0)$. Measurements of the
parity operator $\mathbb{P}$  can be implemented for the
electromagnetic field by using photon counting, or by measuring
the atomic inversion in a micromaser cavity \cite{bge}. For atomic
wave packets or for cold atoms, a parity measurement can be
performed by a measurement of the current position of the particle
relative to a fixed origin \cite{ulm}.

The expectation value of $\mathbb{S}$ is an ``inversion of
probability'' a difference between probabilities of finding a
particle in the positive and negative side of the position axis,
which can be associated with measurements of quadrature
components.

Similarly the expectation value of $\mathbb{R}$ corresponds to an
``inversion of parity" i.e., a  difference between parity
measurements on the positive and negative side of the real axis.
Specific and operational implementations of these measurements for
photons and atoms are under investigation.


\section{Violations of Bell inequalities}
\label{sec6}
\subsection{Displaced $\mathbb{R}$ Operators}
In this section we investigate the violation of the Bell
inequality by the shifted parity inversions $\mathbb{R}$ for Bob
and Alice. We introduce the following correlation function:
\be
E(\alpha,\beta) = \bra{\Psi}\mathbb{R}(\alpha)\otimes
\mathbb{R}(\beta)\ket{\Psi}
\ee

\begin{figure}[h]
\resizebox{0.75\columnwidth}{!}{%
  \includegraphics{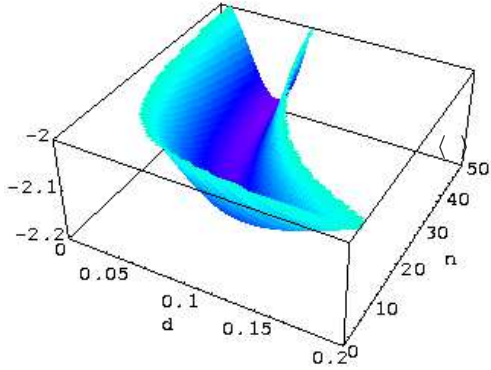}
}
\caption{Plot of the expression $B(d,\sn)$ from Eq. (\ref{bell2a}) for
parity inversions. Only values that exceed the bound imposed by local
theories are shown.} \label{fig2}
\begin{picture}(0,0)(35,10)
\put(205,146){\makebox(0,0){$\lan\;\;\ran$}}
\end{picture}
\end{figure}

\begin{figure}[h]
\resizebox{0.75\columnwidth}{!}{%
 \includegraphics{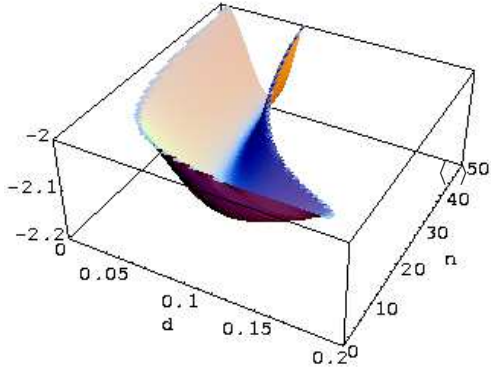}
 }
\caption{Plot of  $B$ from Eq. (\ref{bell2b}). Again, only values that
exceed the bound imposed by local theories are shown.}\label{fig0}
\begin{picture}(0,0)(35,10)
\put(205,125){\makebox(0,0){{\small $\lan\,\,\,\ran$}}}
\end{picture}
\end{figure}

Properly chosen combination of  $ E(\alpha,\beta)$ violates Bell's
inequality. In the  simple case when displacement parameters are
real (e.g. $\A=\mathrm{Re}(\A)=q$ and $\B=\mathrm{Re}(\B)=q'$) the
correlation can be evaluated analytically  and is given by \ba
E(q, q')= \frac{2}{\pi}
{\mathrm{arctan{( 2 \sqrt{\sn (1+\sn)})}}}\;\\
\;\times\,e^{-(  1+2\lan n \ran) \left(q^2+{q'}^2\right)+4
\sqrt{\sn (1+\sn)}qq'} . \n \ea
From this correlation function we form the Bell combination
(\ref{bell1})
\be
\label{bell2a} B(d,\sn)=E(0,0)+E(0,d)+E(-d,0)-E(-d,d)\, ,
\ee
where $d$ and $0$ are the only distance parameters involved in the
 settings. The parameter $\sn$ characterizes the EPR state. The Bell combination
  (\ref{bell2a}) is depicted
 in  Figure \ref{fig2}, and a clear violation of the
bound imposed by local theories can be seen for various values of
$d$ and $\sn$.
If a momentum shift of the parity inversion operators is
performed in addition, no analytical expression for  the correlation can be
obtained. Numerical calculations show, however, that  for
such shifts a violation of Bell's inequality is also possible.
Figure \ref{fig0}
shows expression (\ref{bell1}) for such composed position and
momentum shifts,
\be \label{bell2b}
E(0,0)+E(0,d+i\frac{d}{2})+E(-d+i\frac{d}{2},0)-E(-d+i\frac{d}{2},d+i\frac{d}{2}).\ee
The violation is still noticeable, although for a slightly smaller
range of the shift parameters $d$.

\subsection{Displaced unsharp observables }
\begin{figure}[h]
\resizebox{0.75\columnwidth}{!}{%
  \includegraphics{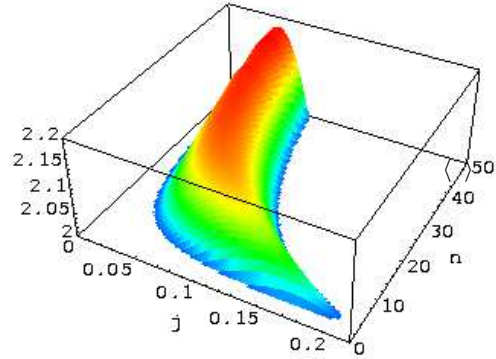}
}
\caption{Combination (\ref{bell1}) as depicted in Figure \ref{fig2} --
except that instead of ${\mathrm{
sgn(q)}}$ we have used here $\tanh(100 q)$ } \label{fig5}
\begin{picture}(0,0)(35,10)
\put(205,124){\makebox(0,0){$\lan\,\,\,\ran$}}
\end{picture}
\end{figure}

When defining the dichotomic operators forming the CV version of
the Pauli algebra we emphasized that it involved an arbitrary
choice. In general, one can introduce an infinite number of
operators with similar properties and probably the only limit
would appear when taking care of their clear physical
interpretation. The parity inversion operator $\mathbb{R}$ was
based on a dichotomic  $\mathrm{sgn}(q)$ function. In this part of
the paper we shall relax this condition and
 instead of the discontinuous
function ${\mathrm{sgn}}(q)$  take a family
 of functions $f_{l}(q,s)$ ($l=1,\, 2,...$) that are not  dichotomic but in some
limit of parameter $s$ represent the sign function. Defining
\ba
\mathbb{R}_l=i\int dq\,  f_l(q,s) |q\ran\lan-q|
 \n
\ea
we obtain operators that may lead to violations of Bell's inequality.
The simplest
examples  of such sequences are  $\,f_1(q,s)=\tanh(sq)\, $ or
\begin{displaymath}
f_2(q,s)=\left\{ \begin{array}{lll} (1-e^{-sq}) & \mathrm{for}& q\geq0\\
(e^{sq}-1)&\mathrm{for}& q\leq0,
\end{array}
\right.
 \end{displaymath}
and
\begin{displaymath}
f_3(q,s)=\left\{ \begin{array}{lll} (1-e^{-sq^2}) & \mathrm{for}& q\geq0\\
(e^{-sq^2}-1)&\mathrm{}& q\leq0.
\end{array}
\right.
 \end{displaymath}
Figure \ref{fig5} reports results obtained for Bell combination
calculated  with $f_1(q,s)=\tanh(sq)$.

\begin{figure}[h]
\resizebox{0.75\columnwidth}{!}{%
  \includegraphics{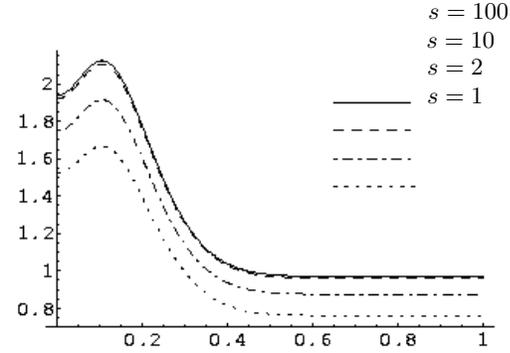}
}
 \caption{Cross-section of Figure \ref{fig5} for $\sn$=10 and values of
parameter $s$
equal to $100$, $10$, $2$, and $1$, respectively.}\label{fig6}
\begin{picture}(0,0)(35,10)
\put(204,142){\makebox(0,0){$s=1$}}
\put(204,153){\makebox(0,0){$s=2$}}
\put(206,163){\makebox(0,0){$s=10$}}
\put(209,174){\makebox(0,0){$s=100$}}
\end{picture}
\end{figure}


\begin{figure}[h]
\resizebox{0.75\columnwidth}{!}{%
  \includegraphics{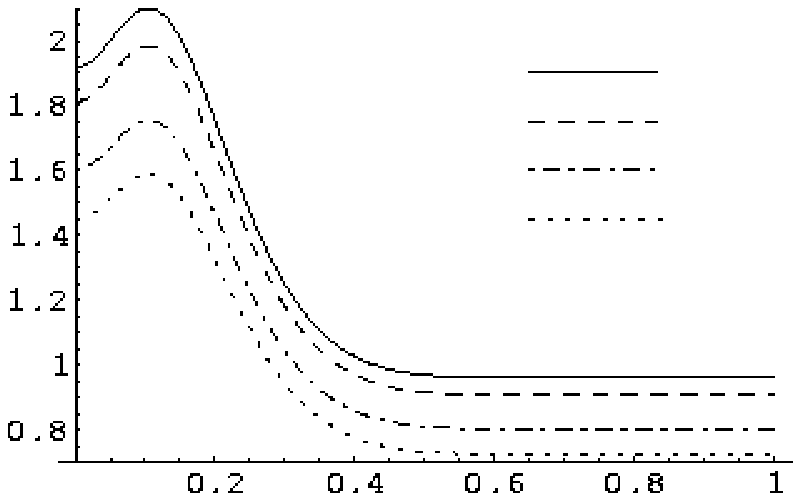}
}
 \caption{Similar cross-section as depicted in Figure \ref{fig6} ($\sn$=10 and
 $s$
equal to $100$, $10$, $2$, and $1$) obtained for $f_3(q,s)$.}
\label{fig8}
\begin{picture}(0,0)(35,10)
\put(204,145){\makebox(0,0){$s=1$}}
\put(204,155){\makebox(0,0){$s=2$}}
\put(206,167){\makebox(0,0){$s=10$}}
\put(209,179){\makebox(0,0){$s=100$}}
\end{picture}
\end{figure}

\noindent The Wigner functions of the corresponding $\mathbb{R}_l$ are of
the form \ba && W_{f_{1}}\sim
\delta(q)\bigg\{{\cal{P}}\big(\frac{1}{p}\big)
-\frac{i}{2s}\bigg[\frac{2s}{p} +\frac{2\pi}{e^{-\frac{\pi
p}{2s}}-e^{\frac{\pi p}{2s}}}
\bigg]\bigg\},\n\\
&& W_{f_{2}}\sim \delta(q)\bigg\{{\cal{P}}\big(\frac{1}{p}\big)
-\frac{s}{s^2+p^2}\bigg\},\n\\
&& W_{f_{3}}\sim \delta(q)\bigg\{{\cal{P}}\big(\frac{1}{p}\big)
-\frac{\sqrt{\pi}}{2s}
\mathrm{erf}\big(\frac{ip}{2s}\big)\bigg\}.\n \ea
In the limit $s\to\infty$  the only terms that do not vanish
are those with $\mathcal{P} \big(\frac{1}{p})$.

An interesting
question is how to determine the smallest value of $s$ sufficient
for a violation of Bell inequality. Figure \ref{fig5} presents
numerical results obtained for $\tanh(100q)$ and Figure \ref{fig6}
shows that although there is no noticeable difference between
$s=100$ and $s=10$, $ s=1$ or $ 2 $ don't lead to functions
changing  rapidly enough near $x=0$ to exceed the bounds imposed by
local theories. Plots obtained for $f_2(q,s)$ do not differ significantly
from that depicted in Figure \ref{fig6}, but analogous plots for
$f_3(q,s)$, Figure \ref{fig8}, show that in this case larger
values of parameter $s$ are needed to provide fully quantum correlations.
This difference is a consequence of the fact that
$\frac{\partial f_3(q,s)}{\partial q}\biggl|_{q=0}=0$

\section{Summary}\label{sum}

We have constructed a class of bounded CV operators and shown that
some of them lead to a violation of Bell's inequality. We have
also provided an explanation based on the Wigner function,  how to
predetermine which operators can lead to such violations.
 This is a general result and one can learn from it at least
two different features: Firstly, as long as the state we
measure/calculate correlations in has a positive Wigner function it
is sufficient to check whether the observables we are interested in
have bounded Wigner functions to decide whether they would
potentially violate
Bell's inequality or not. Secondly, it gives an example of a state
with a positive Wigner function that breaks classical limits and
requires an entirely quantum description.

\section*{Acknowledgments}

This manuscript is based on a talk given at the EU QUEST network
conference {\sl Quantum Information with Photons, and Atoms} (La
Tuile, Italy, March 2004). KW wishes to thank for the kind
hospitality of the National University of Singapore, where this
research has started in the summer of 2003. This work was
partially supported by the Polish KBN grant 2P03 B 02123, 
the European Commission through the Research
Training Network QUEST HPRN-CT-2000-00121, and the Temasek Grant
WBS: R-144-000-071-305.

%
%
%

%

\end{document}